\documentstyle[12pt]{article}
\newcommand{\beeq}{\begin{equation}}
\newcommand{\eneq}{\end{equation}}
\newcommand{\beeqar}{\begin{eqnarray}}
\newcommand{\eneqar}{\end{eqnarray}}
\begin{document}
\vskip 0.5in
\vskip 0.5in
\begin{center}
{\large \bf On Finite Size Effects in $d=2$ Quantum Gravity}\\
\vskip 0.5in
N.D. Hari Dass\\ 
{\bf Institute of Mathematical Sciences, Chennai 600 113, INDIA}\\
\baselineskip 0.5in
{\large \bf Abstract}\\
\end{center}
\indent

A systematic investigation is given of finite size effects in $d=2$ quantum gravity
 or equivalently the theory of dynamically triangulated random surfaces.For Ising
models coupled to random surfaces, finite size effects are studied on the one
hand by numerical generation of the partition function to arbitrary accuracy
by a deterministic calculus, and on the other hand by an analytic theory based
on the singularity analysis of the explicit parametric form of the free energy of the corresponding matrix model.
Both these reveal that the form of the finite size corrections, not surprisingly,
depend on the string susceptibility.For the general case where the parametric form of
the matrix model free energy is not explicitly known, it is shown how to perform the singularity
analysis.All these considerations also apply to other observables like susceptibility
etc.In the case of the Ising model it is shown that the standard Fisher-scaling laws are reproduced.
\vskip 1.0in

A study of finite size effects in statistical systems is of importance for
a variety of reasons. It is indispensable for a more reliable interpretation
of numerical data. As shown by Cardy in the context of systems with conformal
invariance, finite size effects also codify the spectrum of the theory. In most
cases finite size effects are estimated or parametrised by extrapolating the
results of simulations carried out for various sizes. It is the purpose of this
article to show how such finite size effects for two dimensdional
quantum gravity systems can be studied systematically.

Let us begin with the case of so called "pure gravity" which is mapped to the
one-matrix model described by the partition function
\beeq
{\cal Z} = e^{-F} =\int {\cal D}\phi~~~ exp~~ -Tr[{1\over 2}\phi^2 +V(\phi)]
\eneq
where $\phi$ is a $N\times N$ Hermitean matrix. When $V(\phi)= g\phi^4$,the
expression for $F$ in the large-N limit is
\beeq
F = -\sum_n{(-12g)^n(2n-1)!\over n!(n+2)!}
\eneq
interpreting the coefficient of $g^n$ in this sum as the fixed area partition sum for an ensemble
of random surfaces of fixed area $n$,one finds
\beeq
{\cal Z}_n = {(-12g)^n(2n-1)!\over n!(n+2)!}
\eneq
This clearly holds for all $n$.However, we are interested in the thermodynamic or the large-n limit.
To this end we use
\beeq
\Gamma(z)\rightarrow_{z\rightarrow \infty}~~{e^{-z}z^{z-1/2}\over \sqrt{2\pi}}
[1+{1\over 12z}+{1\over 288z^2}+...]
\eneq
for the Gamma-function$\Gamma(z)$.This gives
\beeq
{\cal Z}_n~~=_{n\rightarrow \infty} -{(-1)^n(48)^n~n^{-7/2}\over \sqrt{4\pi}}[1-{25\over 8n}...]
\eneq
The exponential growth in $n$ is non-universal and depends on $V(\phi)$. The
power law correction $n^{-7/2}$ is universal and defines the string susceptiblity
$\gamma$ to be $\gamma=-b+3$ when the power law correction is $n^{-b}$.In this
example the finite size corrections are of the form $[1-{25\over n}+..]$.

Next we consider the case of the Ising model coupled to random surfaces. This is
described by the two matrix model defined by
\beeq
{\cal Z} = e^{-F} =\int {\cal D}\phi_1~{\cal D}\phi_2~~~ exp~~ -Tr[{1\over 2}(\phi_1^2+\phi_2^2)-c\phi_1\phi_2 +(V(\phi_1)+V(\phi_2))]
\eneq
This model is also exactly solvable. But the free energy is not known as an
explicit function of $(c,g)$. However, a parametric form of the free energy
is available in the form
\beeq
F = -{1\over 2}log {g\over z}-{1\over g}\int_0^z~{dt\over t}g(t)+{1\over 2g^2}\int_0^z{dt\over t}g(t)^2
+{1\over 2}log(1-c^2)+{3\over 4}
\eneq
where\beeq
g(z)={z\over (1-3z)^2}-c^2z+3c^2z^3~=~g
\eneq
The numerical generation of ${\cal Z}_n$ is done through the following sequence of steps:i)eliminate
the logarithms in F by considering instead ${\partial F\over \partial g}$, ii)eliminate
inverse powers of $1/(1-3z)$ by repeated use of eqn(). These steps result in a polynomial of degree 8
for ${\partial F\over \partial g}$ with coefficients depending on $(c,g)$. Again
repeated use of eqn() written as a quintic in z results in ${\partial F\over \partial g}$
being expressed as another quintic. Now the numerical technique to solve this
quintic consists of writing
\beeq
z=\sum a_1(n)g^n;~~z^2=\sum a_2(n)g^n;~~.......~~;z^5=\sum a_5(n)g^n
\eneq
Then the quintic(eqn()) is equivalent to the recursion relation
\beeq
a_1(n) = \xi_1a_1(n-1)+\xi_2a_2(n)+\xi_3a_2(n-1)+\xi_4a_3(n)+\xi_5a_4(n)+\xi_6a_5(n)
\eneq
where $\xi_i$ are coefficients depending on $c$.It should be noted that all $a-i(n)$
vanish for negative n and in fact as $z\simeq ~g$ for small g is the branch we 
are interested in, it follows that $a_i(n) = 0$ for $n < i$a and that $a_i(i) = a_1(1)^i$.
It is also easy to see that $a_1(1)=(1-c^2)^{-1}$.

A naive method of solving these recursion relations would make use of the
identities
\beeqar
a_2(n) &=& \sum_1^{n-1}~a_1(m)\cdot a_1(n-m)\nonumber\\
a_3(n) &=& \sum_1^{n-1}~a_1(m)\cdot a_2(n-m)\nonumber\\
 ...      &=& ...\nonumber\\
 ...      &=& ...\nonumber\\
a_5(n) &=& \sum_1^{n-1}~a_1(m)\cdot a_4(n-m)
\eneqar
The computational requirements then grow as $\simeq n^6$ if all $a_i(m);m \le n$
are to be evaluated.This is prohibitive. On the other hand the structure of the eqns(11)
shows that computation of $a_2(n)$ only requires the knowledge of $a_1(m)$ for
$m \le n-1$,computation of $a_3(n)$ requires only the knowledge of$a_2(m)$ for
$m \le n-1$ and hence of $a_1(m)$ for $m \le n-2$ etc.Thus the eqns (10) and (11) can
be solved iteratively simultaneously and the comutational requirement only
grows as $\simeq n^2$ which is entirely manageable.

\noindent
{\bf The Cosmological Constant Problem}\\
\indent
The main difficulty with the abovementioned method is that $a_i(n)$ grow very 
rapidly with n reflecting the (non-universal)exponential growth of ${\cal Z}_n$.For
example,${\cal Z}_{10}\simeq ~ 10^{100}$.In fact by the time one gets to around
100, various coefficients exceed the typical machine limits. Of course, special
purpose arithmetic can be used, but as we shall see soon, a more elegant option
is available. The idea is to go upto some $n_max$ such that the coefficients
are approaching the machine limits and get an estimate for the "cosmological
constant" $\bar\mu$ where ${\cal Z}_n\simeq ~ e^{\bar\mu n}$.Then one scales all
$a_i(n)$ according to
\beeq
\tilde a_i(n) = e^{-\bar\mu n}~~a_i(n)
\eneq
for all i. This results in the scaling
\beeq
\tilde\xi_1~=~e^{-\bar\mu}~\xi_1;~~~~~~~~~\tilde\xi_3~=e^{-\bar\mu}~\xi_3
\eneq
with all other $\xi$'s remaining the same.In the statistical mechanical language this is equivalent to
tuning the system to "criticality".
I will not discuss the results here. They can be seen in \cite{first}.That
reference also carries the details of how a similar analysis can be carried out
for the scaling behaviour of magnetic susceptibility also.

{\bf Singularity analysis of finite size effects}

This is based on the observation that in a sense $n$ and $log g$ are "conjugate" to
each other and that the large-n behaviour of ${\cal Z}_n$ controls the 
convergence of the perturbative expansion in g.In particular this is related to the
behaviour of F near its singular points i.e points where some derivative of F blows up.

Let us illustrate this first with the example of pure gravity.Here the explicit
parametric form of F is
\beeq
F = -{1\over 2}ln z+{1\over 24}(z-1)(9-z)
\eneq
with
\beeq
g(z) = {1-z\over 12z^2} = g
\eneq
The singular point $g_c$ corresponds to $z_0$ where $g^{\prime}(z_0)=0$.Hence
\beeq
z_0=2~~~~~~~~g_c=-1/48
\eneq
Expanding $g(z)$ around $z_0$ one finds
\beeq
g-g_c\simeq~~a(z-z_0)^2+b(z-z_0)^3+...
\eneq
Inverting this one gets
\beeq
z-z_0\simeq~~c(g-g_c)^{1/2}+d(g-g_c)+..
\eneq
Precise values of a,b,c,d can be found in \cite{first}.
It should be noted that $F^{\prime}(z_0)=0$.Expanding F around $z_0$ one
has an expression of the type
\beeq
F(z)-F(z_0)~~=  A(z-z_0)^2+B(z-z_0)^3+...
\eneq
Combining the last eqns one gets for the singular part of the free energy F near $g_c$
the expansion
\beeq
F(g)_{sing}-F(g_c)= {12283\sqrt 3\over 5}(g-g_c)^{5/2}+{1769472\sqrt 3\over 7}(g-g_c)^{7/2}+..
\eneq
It should be remarked that on the basis of eqns(18,19) one would have expected
this expansion to start with $(g-g_c)^{3/2}$.It turns out that the coefficient
of this term vanishes.At this stage it appears accidental, but later when we 
present our improved singularity analysis, we will see that such cancellations
(even more miraculous ones happen in the two-matrix model) have a natural
explanation.
Now we can use the expansion
\beeq
(g-g_c)^{\alpha} = \sum_ng^n~(-g_c)^{\alpha-n}{\Gamma(\alpha+1)\over
\Gamma(n+1)\Gamma(\alpha-n+1)}
\eneq
and the asymptotic expansion of Gamma functions(eqn()) to get the asymptotic
expansion for ${\cal G}_n$ defined by
\beeq
(g-g_c)^{\alpha}=\sum_n g^n~{\cal G}_n
\eneq
giving
\beeq
{{\cal G}_n}_{n\rightarrow \infty}~~=(-1)^{n+1}(-g_c)^{\alpha-n}~\Gamma(\alpha+1)~n^{-(1+\alpha)}
{\sin \pi\alpha\over \pi}e^{\alpha(1+\alpha)/2n}
\eneq
Combining eqns(19,21,22) yields eqn(5).

In the case of the two-matrix model, the singularity analysis is a bit more 
involved.In the case of the $\phi^4$-potential,$g^{\prime}(z_0)=0$ yields
$z_0=-1/3$ in the low temperature phase and a known function $z_0(c)$ in the
high temperature phase.From the parametric representation it is easy to see that
$F^{\prime}(z_0)=0$ always. Boulatov and Kazakov \cite{bk} had claimed that
$F^{\prime\prime}(z_0)$ is also zero.But we find that this is so only at the
Ising critical point $c=c_{cr}$.Now we present the singularity analysis for the cases
$c\ne c_{cr}$ and $c=c_{cr}$ separately.

{\bf Non-critical Case}\\
Here the situation is identical to the pure gravity case except that all the coefficients
in the expansion depend on c.Again, the coefficient of $(g-g_c)^{3/2}$ in F,
which is now a function of c, vanishes for all c!Apart from reproducing the
correct string susceptibility, the form of the finite size corrections one
obtains are:
\beeqar
High Temp Phase(c=.36) &:&  ~~~~1-{8.76\over n}+...\nonumber\\
Low Temp Phase(c=.20) &:&  ~~~~1-{72.69\over n}+...
\eneqar
{\bf Critical Case}\\
In this case we have $F^{\prime\prime}(z_0)=0,g^{\prime\prime}(z_0)=0$
in addition to the vanishing of the corresponding first derivatives.Thus
eqns(17-19) are now replaced by
\beeq
g-g_c\simeq~~a(z-z_0)^3+b(z-z_0)^4+...
\eneq
\beeq
z-z_0\simeq~~c(g-g_c)^{1/3}+d(g-g_c)^{2/3}+h(g-g_c)+..
\eneq
\beeq
F(z)-F(z_0)~~=  A(z-z_0)^3+B(z-z_0)^4+...
\eneq
Again the $(g-g_c)^{4/3},(g-g_c)^{5/3}$ terms in the singular part of the
free energy drop out and the leading singular part $\simeq (g-g_c)^{7/3}$
characterstic of the string exponent$\gamma = -1/3$.The form of the finite
size corrections are
\beeqar
V(\phi) &=&~\phi^4 : ~~~~1+{0.4287\over n^{1/3}}-{3.01\over n}-{1.298\over n^{4/3}}...\nonumber\\
V(\phi) &=&~\phi^3 : ~~~~1+{0.2860\over n^{1/3}}-{3.05\over n}-{0.934\over n^{4/3}}...
\eneqar
These eqns represent the most important results of our analysis,namely,that
the form of the finite size corrections is system-dependent and in this particular
example depend on the string susceptibility $\gamma$.It should be noted that in
eqn(28) there are no terms of the type $n^{-2/3}$. This will become clearer with our improved singularity analysis.
It should also be noted from eqn(28) that the coefficients in the finite size factor are non-universal.
This means it is possible to choose an appropriate potential that will minimise
the finite size corrections. This situation is well known in the study of lattice
gauge theories("improved action"-principle). 

On the basis of eqn(28) it had been conjectured in \cite{first} that if the
string susceptibility is of the form $\gamma=p/q$ with p,q relatively prime,the
form of the finite size corrections will be
$1+{a\over n^{1/q}}+...$. We will take up this issue now.

{\bf Improved Singularity Analysis}

The singularity analysis presented above relied on the availability of the
explicit form of the parametrised free energy. In most cases such explicit
forms are not available and the above-presented method will fail in those cases.
There are many models of interest like multi-critical matrix models, models
with $\gamma > 0$ \cite{sdas} whose finite size analysis one may wish to
perform.

In all such cases where the method of orthogonal polynomials \cite{mehta} can be applied,
the free energy is erxpressible as
\beeq
F~=~\int_0^1~d\xi~(1-\xi)~ln~f(\xi)
\eneq
where the function $f(\xi)$ is defined through the method of orthogonal polynomials
\beeqar
\int~dx~w(x)P_i(x)P_j(x)~&=&~h_i\delta_{ij}~~~~~~~One Matrix Models\nonumber\\
\int~dx~dy~w(x,y)P_i(x)P_j(y)~&=&~h_i\delta_{ij}~~~~~~~Two Matrix Models
\eneqar
\beeq
f_i~=h_i/h_{i-1}~\rightarrow_{N\rightarrow \infty}~~f(\xi)
\eneq
In what follows we shall only consider one matrix models but consider arbitrary potentials.
The analog of the defining eqns for $g(z)$ is now
\beeq
g\xi = w(f(\xi))
\eneq
In all these eqns the explicit form of $w(x)$ is dictated by the potential.
The multicritical points are determined by the vanishing of various derivatives of $w(x)$.
The upshot of all this is that
multicritical models are characterised by
\beeq
g\xi-g_c~=~A(f(\xi)-B)^p+C(f(\xi)-D)^{p+1}+..
\eneq
The meaning of this equation is that all parameters but $g$ have been fixed at their critical
values and $g$ is near $g_c$.Further, if $w(x)$ is chosen to be the smallest polynomial
satisfying multicriticality, there is only one term in eqn(38).

The parametric representation given earlier in terms of the variable $z$
can be recovered on identifying $f(1)$ with $z$ whence eqn(37) reads as $g=w(z)= g(z)$.
Inverting eqn(38)
\beeq
f(\xi) = B^{\prime}~+A^{\prime}(g\xi-g_c)^{1/p}+..
\eneq
On using this with eqn(29) one finds
\beeq
F_{sing}~\simeq~~\sum_{m=1}^{m=p-1}~(g-g_c)^{2+m/p}
\eneq
A naive expectation for the exponent would be $1+m/p$, but the $(1-\xi)$ factor in the
integrand suppresses this dominant exponent.This is basically what was behind the surprising cancellations we alluded
to earlier.

Thus we see that in the fractional powers of $1/n$ characterising finite size corrections,
two terms will be missing,one corresponding to the first term in eqn(35) which
controls the leading behaviour and one corresponding to the term $m=p$ which
is regular. Now we quote results for various cases:
\begin{center}
%\begin{table}
\begin{tabular}{lll}
\hline
\hline\\
$~~~n~~~$ & $~~~\gamma~~~$& Finite Size Correction  \\ 
\hline
\hline\\
 ~~~~~2 &~~~$-1/2$  &~~~$ 1+{a_1\over n}+{a_2\over {n^2}}+...     $\\ \\ 
 ~~~~~3 &~~~$-1/3$  &~~~$ 1+{a_1\over {n^{1/3}}}+{a_2\over {n^{4/3}}}+... $    \\  \\
 ~~~~~4 &~~~$-1/4$  &~~~$ 1+{a_1\over {n^{1/4}}}+{a_2\over {n^{2/4}}}+... $    \\  \\
\hline
\hline\\
\end{tabular}
%\end{table}
\end{center}
Likewise, the results for models with positive $\gamma$ \cite{sdas} are
given below: 
\begin{center}
%\begin{table}
\begin{tabular}{lll}
\hline
\hline\\
$~~~m~~~$ & $~~~\gamma~~~$& Finite Size Correction  \\ 
\hline
\hline\\
 ~~~~~2 &~~~1/2  &~~~$ 1+{a_1\over n}+{a_2\over {n^2}}+...  $   \\  \\
 ~~~~~3 &~~~1/3  &~~~$ 1+{a_1\over {n^{2/3}}}+{a_2\over {n^{5/3}}}+...$     \\  \\
 ~~~~~4 &~~~1/4  &~~~$ 1+{a_1\over {n^{2/4}}}+{a_2\over {n^{3/4}}}+...$     \\  \\
\hline
\hline\\
\end{tabular}
%\end{table}
\end{center}

Indeed there are many interesting issues that need to be properly understood in this context.
For example, in the continuum version of Distler and Kawai \cite{DK}, and of
David \cite {D}, the liouville field is integrated over all its possible values. In such a
treatment no "finite size" corrections will ever appear. The fixed area formulae hold for all areas.It would be interesting to
formulate the problem of scaling violations in such approaches.Other issues of both practical and theoretical
interest are the finite size analysis for other observables like the Hausdorff dimension, loop
length distributions, resistivity of random networks etc. Another outstanding issue is
to understand what happens in the $c=1$ case where $\gamma=0$ and there are 
logarthmic modifications to the 
fixed area partition sum.

\end{document}